\documentclass[8.5pt,twoside,twocolumn]{article}
\usepackage[super,sort&compress,comma]{natbib} 
\usepackage{mhchem}
\usepackage{times,mathptmx}
\usepackage{sectsty}
\usepackage{balance} 
\usepackage{graphicx}
\usepackage{lastpage}
\usepackage[format=plain,justification=raggedright,singlelinecheck=false,font=small,labelfont=bf,labelsep=space]{caption} 
\usepackage{amsmath}
\usepackage{amssymb}
\usepackage{upgreek}
\usepackage{dblfloatfix}
\newcommand{\celc}{\ensuremath{\,^\circ\textrm{C}}}
\newcommand{\sti}{\text{type-M}}
\newcommand{\stii}{\text{type-B}}
\newcommand{\stiii}{\text{type-I}}
\DeclareMathAlphabet{\mathcal}{OMS}{cmsy}{m}{n}
\newcommand{\RR}{\ensuremath{\mathcal{R}}}

\usepackage{fancyhdr}
\begin{document}
\thispagestyle{plain}
\fancypagestyle{plain}{
\renewcommand{\headrulewidth}{1pt}}
\renewcommand{\thefootnote}{\fnsymbol{footnote}}
\renewcommand\footnoterule{\vspace*{1pt}%
\hrule width 3.4in height 0.4pt \vspace*{5pt}} 
\setcounter{secnumdepth}{5}
\makeatletter 
\def\subsubsection{\@startsection{subsubsection}{3}{10pt}{-1.25ex plus -1ex minus -.1ex}{0ex plus 0ex}{\normalsize\bf}} 
\def\paragraph{\@startsection{paragraph}{4}{10pt}{-1.25ex plus -1ex minus -.1ex}{0ex plus 0ex}{\normalsize\textit}} 
\renewcommand\@biblabel[1]{#1}            
\renewcommand\@makefntext[1]%
{\noindent\makebox[0pt][r]{\@thefnmark\,}#1}
\makeatother 
\renewcommand{\figurename}{\small{Fig.}~}
\sectionfont{\large}
\subsectionfont{\normalsize} 

\renewcommand{\headrulewidth}{1pt} 
\renewcommand{\footrulewidth}{1pt}
\setlength{\arrayrulewidth}{1pt}
\setlength{\columnsep}{6.5mm}
\setlength\bibsep{1pt}

\twocolumn[
  \begin{@twocolumnfalse}
\noindent\LARGE{\textbf{Capillary leveling of stepped films with inhomogeneous molecular mobility}}
\vspace{0.6cm}

\noindent\large{\textbf{Joshua D. McGraw,\textit{$^{ac}$} Thomas Salez,\textit{$^{b}$}
Oliver B\"aumchen,\textit{$^{a}$} \'Elie Rapha\"el,\textit{$^{b}$} and Kari Dalnoki-Veress\textit{$^{ab\ast}$}}}\vspace{0.5cm}

\noindent \normalsize{A homogeneous thin polymer film with a stepped height profile levels due to the presence of Laplace pressure gradients. Here we report on studies of polymeric samples with precisely controlled, spatially inhomogeneous molecular weight distributions. The viscosity of a polymer melt strongly depends on the chain length distribution; thus, we learn about thin-film hydrodynamics with viscosity gradients. These gradients are achieved by stacking two films with different molecular weights atop one another. After a sufficient time these samples can be well described as having one dimensional viscosity gradients in the plane of the film, with a uniform viscosity normal to the film. We develop a hydrodynamic model that accurately predicts the shape of the experimentally observed self-similar profiles. The model allows for the extraction of a capillary velocity, the ratio of the surface tension and the viscosity, in the system. The results are in excellent agreement with capillary velocity measurements of uniform mono- and bi-disperse stepped films and are consistent with bulk polymer rheology. }

\vspace{0.5cm}
 \end{@twocolumnfalse}
  ]
\section{Introduction}
\footnotetext{\textit{$^{a}$~Department of Physics \& Astronomy and the Brockhouse Institute for Materials Research, McMaster University, Hamilton, ON, Canada. E-mail: dalnoki@mcmaster.ca}}
\footnotetext{\textit{$^{b}$~Laboratoire de Physico-Chimie Th\'eorique, UMR CNRS Gulliver 7083, ESPCI, Paris, France. }}
\footnotetext{\textit{$^{c}$~Present address: Department of Experimental Physics, Saarland University, D-66041 Saarbr\"ucken, Germany.}}
\label{intro}
The properties of nanoscale polymer systems have received considerable recent attention~\cite{tsui08TXT}. This interest is partly due to the fact that polymer films have enormous technological importance, such as for the fabrication of computing components~\cite{jaeger02TXT, tang08SCI, park09SCI}. Thin polymer-containing membranes are also vital in many biological systems~\cite{craster09RMP, eijkel05MNF}. More generally, thin polymer films are attractive systems for the study of confined molecules because they can be easily prepared, for example by spin-coating from a volatile solvent, such that the resulting sample has a thickness that is comparable to the typical size of molecules (some tens of nanometers). Finally, the macromolecular nature of polymer chains allows one to make quantitative statements about the observed behaviours without much regard for the specific chemical makeup of the monomer units~\cite{degennesscaling, matsenJPCM02}.

Confined polymeric systems can exhibit anomalous dynamics~\cite{fakhraai08SCI, paeng11JACS}. In the last two decades, several researchers~\cite{keddie94EPL, forrest01ACIS, seemann06JPS, ellisonNATM03} have observed that the glass transition temperature, $T_g$, of polymer films is reduced when the thickness of the films is below some critical value. A reduced glass transition may be indicative of a faster dynamics even above $T_g$; therefore, a film with thickness gradients may experience laterally inhomogeneous dynamics. Similarly, for polymer melts, several groups~\cite{allan, oconnellSCI2006, Bodiguel2006, shinNATM07} have reported an enhanced dynamics or an altered entanglement network when a system dimension is less than the typical polymer coil size. As in the case of $T_g$ reductions, a liquid film with height gradients may experience inhomogeneous dynamics if there are regions with thicknesses comparable to the size of the molecules. 

Given that there is such great interest in the topic of nanoscopic polymer systems, it has become necessary to find new tools for the investigation of polymer properties on small scales. In the last couple of decades, several such tools have been developed. Because of the large surface to volume ratio in comparison to bulk systems, many of the developed techniques take advantage of surface or interfacial tension~\cite{isf11TXT}. In the case of dewetting of polymer films~\cite{reiter92PRL, reiter08TXT, seemann01PRL, seeman01PRLb}, holes grow in order to expose a low surface energy substrate, thereby reducing the free energy of the system. The observation of hole growth in free standing films~\cite{dalnoki01PRE, roth06JPSB} is a similar scenario that can also be used to study high strain rate nanoscopic flows. Another example for which surface tension is the driving force is the flow of a polymer melt into nanoporous alumina~\cite{shinNATM07}. In experiments for which the surface tension does not provide the sole driving force, the use of electrohydrodynamic instabilities~\cite{barbero2009, thomas11PRE, Closa2011} has contributed to the understanding of some non-equilibrium properties of thin polymer films. Additionally, the use of a micro-bubble inflation technique~\cite{oconnellSCI2006} has contributed to the study of polymer films with thickness comparable to molecular size. Here we continue our investigations of nanometric polymer systems in which a surface tension driven flow is observable. The samples studied serve as model systems through which one may understand the more complicated dynamics in confined systems. 
\begin{figure}[t!]
	\centering 
	\includegraphics{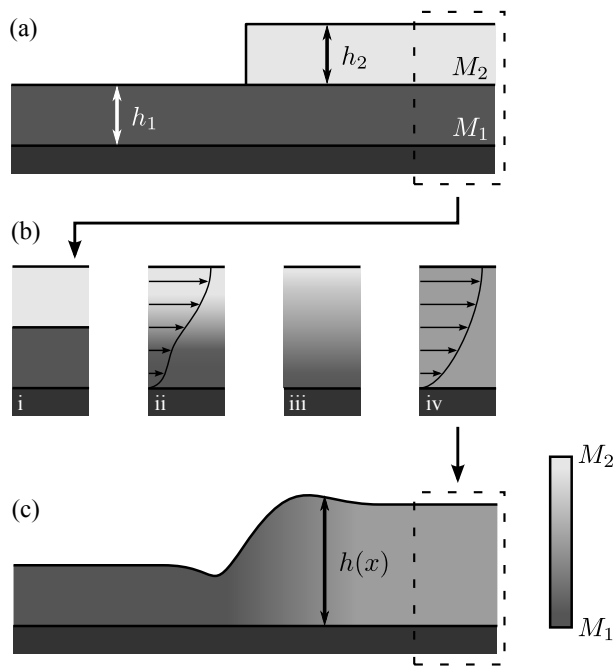}
	\caption{A schematic showing (a) as-prepared inhomogeneous samples with composition 
$\{M_1, M_2\}$ and geometry $\{h_1, h_2\}$; preparation details may be found in Section~\ref{exp}. (b) Evolution of the vertical composition profile. The transient regime in i - iii is characterized by a non-constant vertical molecular weight distribution. In ii, the arrows schematically represent the resulting velocity profile if $M_1 > M_2$ in the presence of a lateral pressure gradient. In iv, the film is vertically homogenized, and a parabolic flow profile (arrows) can be assumed. The transient regime occurs on timescales that are short compared to the accessed experimental times, and so is not discussed in detail in this article. (c) Composition and height profiles after annealing for a period of time sufficient to reach the self-similar regime (described in Section~\ref{ssim}). The greyscale intensity schematically indicates the weight-averaged molecular weight. In the text, these are referred to as \stiii\ samples. } 
	\label{sampschem} 
\end{figure} 

In previous studies~\cite{mcgraw11SM, mcgraw12PRL, salez12EPJE, salez12PoF} we investigated flows in leveling stepped films with no mobility gradients. Fig.~\ref{sampschem}(a) with $M_1 = M_2$, where $M_i$ is the molecular weight of the polymers making up a film with height $h_i$, schematically shows the samples we studied in those works. We have demonstrated that for nanometric films whose thickness is much greater than the unperturbed polymer size, the evolution is well described by a lubrication approximation of the Stokes equations~\cite{oron97RMP, craster09RMP}. We found that the evolution of the film height profile with time is governed by a balance between a capillary driving force and a viscous friction. In agreement with the theory, the experimental profiles were found to be self-similar in the variable $u = xt^{-1/4}$ where $x$ is the lateral position, and $t$ is the time~\cite{mcgraw12PRL, salez12EPJE, salez12PoF}. The form of this variable comes about naturally from the dimensionality of the governing equation~\cite{stillwagonJAP88}, which has a first order temporal derivative and a fourth order spatial derivative, see Eqs.~(\ref{hxt}) and~(\ref{xX}). In this work, we will also make significant use of this self-similarity variable. 

In the present work, we demonstrate that stepped films can be instructive model systems for the study of thin, inhomogeneous fluids. The samples at the focus of this study are schematically shown in Fig.~\ref{sampschem}. They are prepared using polymers with two different molecular weights. Samples are prepared such that the molecular weight distribution remains inhomogeneous throughout the observed evolution. The mobility of polymer liquids is highly dependent on their chain length~\cite{struglinski85MAC, kramer84POL, greenMAC86, liuMAC93, karimMAC94, pierce11EPL, wang11JoR}. As such, a deviation from the behaviour observed in homogeneous systems is expected because the mobility in the sample shown in Fig.~\ref{sampschem} is not uniform. This spatial inhomogeneity of the chain length distribution deliberately violates one of the assumptions of the model used to describe the samples in our previous studies~\cite{mcgraw11SM, mcgraw12PRL, salez12EPJE, salez12PoF}. By making comparisons between the profiles of inhomogeneous samples, and geometrically similar (\textit{i.e.} similar $\{h_1,  h_2\}$) homogeneous samples, we will see that identifying deviations from bulk behaviour may be possible. 

This communication is organized as follows. In Section~\ref{exp}, we describe in detail the sample preparation and the experimental procedures used. Section~\ref{resmod} is dedicated to a qualitative description of the experimental results and to a presentation of the models used to describe these results. The hydrodynamic model used to describe the inhomogeneous samples admits a self-similar solution that is observed in the experiments. In Section~\ref{discuss}, we present a discussion of the capillary velocities as measured in homogeneous samples, and then turn to an analysis of the self-similar evolution of the inhomogeneous samples. 

\section{Experiment}
\label{exp}
The samples described in this work were all prepared from polystyrene (PS) purchased from Polymer Source Inc. Weight-averaged molecular weights, $M$, ranged from 15.5 to 592 kg/mol and had polydispersity indices, $\rm{PI} < 1.1$. See Table~\ref{ps} for details. PS was dissolved into toluene (Fisher Scientific, Optima grade) in various weight fractions ranging from 0.5 -- 6 wt\%. In this study, two types of solutions were prepared. First, one component solutions were prepared in which PS of a single molecular weight was dissolved in toluene. Second, two component solutions were prepared into which various weight fractions, $\phi$, of a given molecular weight were dissolved in toluene. In both cases, films with thickness $h_1$ were prepared by spin-coating the polymer solution onto Si wafers (University Wafer) rinsed with ultra pure water (18 M$\Upomega$\,cm, Pall Cascada, LS), methanol (Fisher Scientific, Optima grade) and toluene. Additionally, films with thickness $h_2$ were spin-coated onto freshly cleaved mica (Ted Pella Inc.) substrates. Film thicknesses were always much larger than the unperturbed radii of gyration~\cite{cotton74MAC}, and were never larger than $\sim$\,200 nm. All films on Si and mica were pre-annealed at 130\celc\ for 24 hr in a home-built vacuum oven prior to stepped film preparation in order to remove residual solvent and to allow relaxation of the polymer chains. 
\begin{figure}[t!]
	\centering 
	\includegraphics{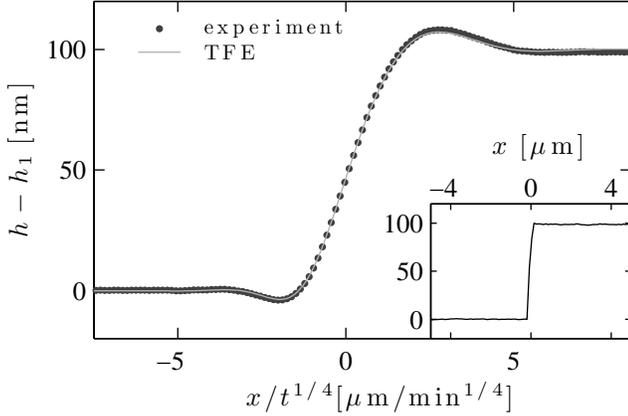}
	\caption{ A monodisperse 15.5\ kg/mol stepped film annealed for 10\ min at 140\celc, with heights $\{h_1, h_2\} = \{102, 98\}$\ nm. Data points represent AFM height data while the grey solid line represents a uniformly scaled numerical solution of TFE (see Eq.~(\ref{hxt})) with a Heaviside initial condition~\cite{salez12EPJE}. Inset: the experimental height profile before annealing. }
	\label{typei} 
\end{figure} 

\begin{table}[b!]
\centering
  \caption{Weight-averaged molecular weights  (in kg/mol) and polydispersity indices (PI) of all PS used for the sample preparation in this study.}
\begin{tabular}{l r r r r r}
  \hline
  $M$ & 15.5 & 24.7 & 31.8 & 55.5 & 118 \\
  PI  & 1.04 & 1.03 & 1.06 & 1.07 & 1.05 \\
  \hline
  \end{tabular}
  \label{ps}
  \begin{tabular}{l r r r r}
   \hline
  $M$ & 192 & 286 & 373 & 592 \\
    PI & 1.04 & 1.06 & 1.07 & 1.09 \\
  \hline  
\end{tabular}
\end{table}
After pre-annealing the individual films above the glass transition temperature, $T_g\approx 100\celc$, the stepped films depicted schematically in Fig.~\ref{sampschem} were prepared as detailed elsewhere~\cite{mcgraw11SM}. Briefly, PS films on mica were floated onto the surface of an ultra-clean water bath, and transferred onto a plasma cleaned (air, 30\ s on low power, Harrick Plasma) Si wafer. The films were allowed to dry and a diamond scribe was pressed firmly onto the Si wafer causing the wafer to break along a crystal plane. This breaking also splits the nanometric PS film, producing a clean straight edge along the PS film that is free of crazes~\cite{kramercrrev2} in some regions. The films on the cracked Si wafers were floated back onto the surface of a water bath, and were subsequently picked up with the film cast on Si. The floating process creates a stepped film of lower height $h_1$ and  step height $h_2$ -- the inset of Fig.~\ref{typei} shows an initial height profile as measured with atomic force microscopy (AFM) at room temperature. Three types of stepped films were prepared:  
\begin{itemize}
\item \underline{\sti} (homogeneous, monodisperse $M$ distribution): samples containing PS of a single molecular weight, $M$, were prepared to measure the molecular weight dependence of the viscosity in homogeneous samples. The molecular weights used for homogeneous samples are shown in Table~\ref{ps}.
\item \underline{\stii} (homogeneous, bidisperse $M$ distribution): stepped films whose two layers were cast from solutions containing a blend of two molecular weights. The resulting spin-cast films had a weight fraction $\phi$ of PS with molecular weight $M_l$ and weight fraction $1-\phi$ of PS with molecular weight $M_s$, where the subscript`$l$' refers to the larger molecular weight and the subscript `$s$' refers to the smaller one. The composition of these samples is given as $\{\phi; M_l, M_s\}$. In this study, we have used the combinations $\{\phi; M_l, M_s\} = \{\phi; 192, 55.5\}$, $\{\phi; 55.5, 15.5\}$ and $\{\phi; 118, 15.5\}$ with various $\phi$, and with the molecular weights in kg/mol.
\item \underline{\stiii} (laterally inhomogeneous, vertically bidisperse $M$ distribution): stepped films whose component films were made from two different chain length polymers. A film with molecular weight $M_1$ cast on Si, and a film with molecular weight $M_2$ cast onto mica. The film on mica was then floated onto the film on Si. These samples are schematically shown in Fig.~\ref{sampschem} and are referred to as having composition $\{M_1, M_2\}$. In this study, we have used the combinations $\{M_1, M_2\} = \{55.5, 192\}, \{192, 55.5\} , \{15.5, 118\}$ and $\{118, 15.5\}$\ kg/mol.
\end{itemize}
\begin{figure*}[t!]
  \centering
\includegraphics{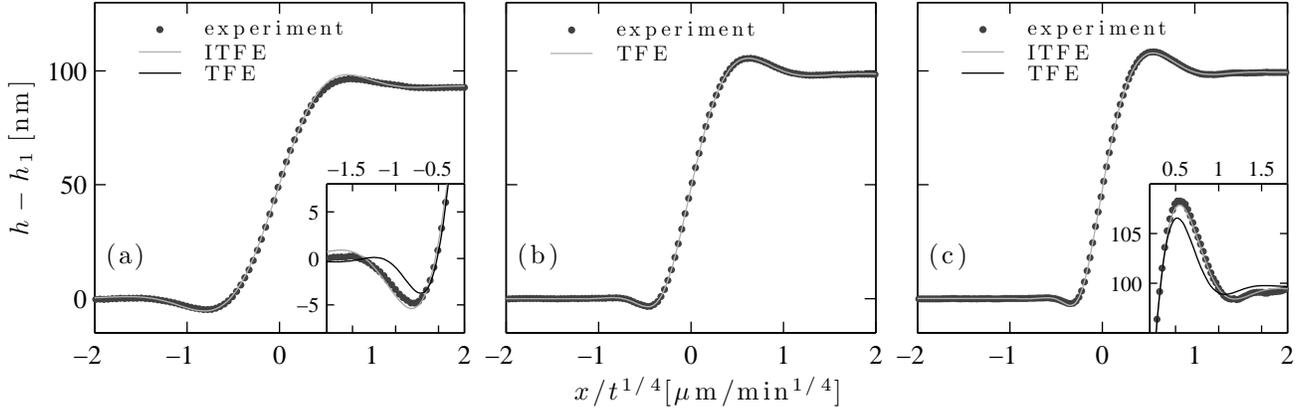}
  \caption{ Height as a function of the scaled position for bidisperse (\stii) and inhomogeneous (\stiii) samples. Data points in all graphs are from AFM profiles.  All samples represented in this figure were measured after 240 min of annealing at 140\celc. (a) A \stiii\ sample that was prepared according to Fig.~\ref{sampschem} with $\{M_1, M_2\} = \{55.5, 192\} $\ kg/mol and $\{h_1, h_2\} = \{101, 93\}$\ nm. The grey line represents the best fit to a numerical solution of ITFE (see Eq.~(\ref{ITFEnd})). (b) A \stii\ sample with $\{\phi; M_l, M_s\} = \{0.50; 192, 55.5\}$ and $\{h_1, h_2\} = \{99, 99\}$\ nm, where the molecular weights are in kg/mol. The grey line represents the best fit to a numerical solution of TFE (see Eq.~(\ref{hxt})). (c) A \stiii\ sample with $\{M_1, M_2\} = \{192, 55.5\}$\ kg/mol and $\{h_1, h_2\} = \{107, 99\}$\ nm -- the grey line is a best fit to an ITFE profile. In (a) and (c), the insets highlight the dip and bump regions, respectively; the black line is a best fitting profile computed according to Eq.~(\ref{hxt}) and the grey line is a best fit profile computed according to Eq.~(\ref{ITFEnd}).}
\label{animals}
\end{figure*}

With stepped films prepared, we measured $h_2$ using AFM. The samples were then heated to $140\celc > T_g$ for which a capillary-driven viscous flow is induced by the excess surface of the height profile. The stepped films were observed for times $10 < t < 51\,500$\ min. The large range of molecular weights used requires a large range of times to ensure self-similarity of the profiles. After various evolution times, the samples were quenched to room temperature and height profiles were measured using AFM. We stress the fact that flow proceeds \emph{above} the glass transition, but that our AFM measurements have been performed after a quench to room temperature, which is $\sim80$\,\celc\ \emph{below} the glass transition. An example of a \sti\ sample is shown in Fig.~\ref{typei}, and examples for \stii\ and \stiii\ samples are shown in Fig.~\ref{animals}. After self-similarity in $xt^{-1/4}$ was verified by evaluating the profiles for at least two different times, a scratch down to the Si substrate was made in the film using a scalpel blade and $h_1$ was then measured with AFM. 
  
\section{Results and models}
\label{resmod}
This section is divided as follows. In part~\ref{gogc}, we give a description of the results. We then outline the model used to describe monodisperse and bidisperse samples in part~\ref{t1mod}, followed by part~\ref{modA} where we describe the model for inhomogeneous samples. 
\subsection{General overview of profile shapes}
\label{gogc}

In Fig.~\ref{typei} is shown the result of a leveling experiment for a monodisperse (\sti) sample with $M=15.5$~kg/mol. The height data was obtained after the as-prepared stepped film was annealed for 10 min at 140\celc. This type of sample was examined in our previous studies~\cite{mcgraw11SM, mcgraw12PRL, salez12EPJE} and is used here to establish the scaling of inverse capillary velocity, $\eta/\gamma$, with molecular weight as described in part~\ref{cvos}. Here, $\gamma$ is the liquid-air surface tension, and $\eta$ is the viscosity. In Fig.~\ref{animals}, we show height profiles of  \stii\ and \stiii\ samples that have been annealed for 240\ min at 140\celc. For each profile shown in Figs.~\ref{typei} and~\ref{animals}, the heights are all approximately the same: $h_1\approx h_2 \approx 100$\ nm. Since this is the case, the differences observed in comparing the profile shapes in Figs.~\ref{animals}(a) and~\ref{animals}(c), to that shown in Fig.~\ref{animals}(b) are dominated by the differences in the molecular weight distribution of the samples. 

One way of characterizing stepped film height profiles is through the ratio:
\begin{equation}
\RR \equiv Y_b/Y_d\ , 
\label{bddef}
\end{equation}
where we define the `bump' and `dip' heights as $Y_b=h_\textrm{max}-(h_1+h_2)$ and $Y_d = h_1 - h_\textrm{min}$. Here, $h_\textrm{max}$ and $h_\textrm{min}$ refer to the global maximum and minimum of the height profile. A large value of \RR\ is indicative of a high mobility on the high side of the stepped film. For 11 homogeneous \sti\ samples with $M = 15.5$\ kg/mol over a range of geometries,  $1.2 < h_1/h_2 < 7.4$, we find that the experimental data satisfies the empirically determined relation:
\begin{equation}
\RR = \frac{h_1+h_2}{h_1}~~\textrm{[homogeneous]}\ ,
\label{bd}
\end{equation}
which is in agreement with data from computational simulations~\cite{salez12EPJE} over a similar range of height ratios.

\begin{figure}[t!]
	\centering 
	\includegraphics{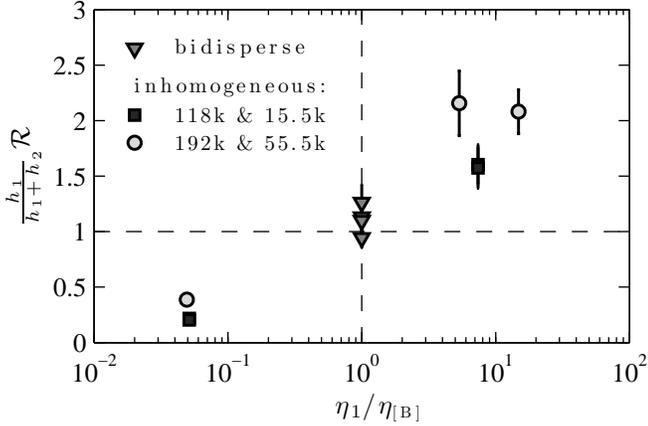}
	\caption{\RR\ from Eq.~(\ref{bddef}), normalized by the ratio of heights for an as-prepared stepped film, as a function of $\eta_1/\eta_\textrm{[B]}$. $\eta_1$ is the viscosity of the molecular weight comprising the lower layer of the as-prepared sample. $\eta_{[\textrm{B}]}$ is the viscosity of a bidisperse sample whose composition is commensurate with the height ratio of the corresponding \stiii\ sample.} 	\label{bdplot} 
\end{figure}

The data in Figs.~\ref{animals}(a) and \ref{animals}(c) show \RR\ values that are respectively smaller and bigger than the \RR\ for the homogeneous sample in Fig.~\ref{animals}(b). In Fig.~\ref{bdplot}, we plot \RR, normalized according to Eq.~(\ref{bd}), as a function of the viscosity ratio, $\eta_1/\eta_\textrm{[B]}$, for all the \stiii\ samples studied here, and for several \stii\ samples (\emph{n.b.} for \stii\ samples, $\eta_1 = \eta_\textrm{[B]}$). $\eta_1$ is the viscosity\footnote{See part~\ref{t1mod} for a description the capillary velocity, $\gamma/\eta$, determination. We also assume that the surface tension, $\gamma$, is constant. } of the molecular weight comprising the lower layer of the as-prepared sample. $\eta_{[\textrm{B}]}$ is the viscosity of a bidisperse sample whose composition is commensurate with the height ratio of the corresponding \stiii\ sample. We note that the data are always above unity on the vertical coordinate for $\eta_1/\eta_\textrm{[B]} > 1$, while the data are always below unity for $\eta_1/\eta_\textrm{[B]} < 1$. 

The results of Figs.~\ref{animals}(a) and~\ref{animals}(c) and Fig.~\ref{bdplot} can be described in terms of the mobility distribution of the observed samples. In the initial state, all of the excess free energy is localized in the height step at $x = 0$. When the sample is heated from room temperature to above the glass transition, molecules move in response to gradients in the Laplace pressure at the fluid-air interface, which results from the gradients in the curvature of the free surface~\cite{degennes03TXT}. If $\gamma$ is the surface tension,  then we can approximate the pressure by $p \approx -\gamma \partial_x^2 h$ when the height variations are small~\footnote{Though strictly not the case at $t=0$ in our experiments, the lubrication approximation is valid for the detailed comparisons we make between theory and experiment, after the films have leveled significantly.}. Since the geometries of the samples in Fig.~\ref{animals} are nearly identical, the driving force at $t=0$ is the same for all samples, and the local flux is set by the local viscosity only: a low viscosity gives rise to larger volume transport. Therefore, a low viscosity region on the thin side of the stepped film gives rise to a smaller \RR\ (larger $Y_d$) compared to a homogeneous sample with identical geometry. Conversely, if the viscosity on the thin side of the stepped film is higher than the viscosity on the thick side, we expect flow to be favoured on the thick side, thus we expect \RR\ to be larger (larger $Y_b$) than for the homogeneous case. The qualitative features of each \RR\ in Figs.~\ref{animals}(a) and~\ref{animals}(c), and the trend observed in Fig.~\ref{bdplot}, are consistent with this picture. 

We finally note that the slopes of the profiles at $xt^{-1/4} = 0$ are increasing monotonically from left to right in Fig.~\ref{animals}, with a higher slope indicating a slower evolution. In each case, the composition of the samples on the high side of the stepped films is approximately equal. However, the molecular weight of the samples' thin sides is increasing from left to right in Fig.~\ref{animals}. Thus, as the viscosity on the thin side of the stepped film increases, so too does the slope at the midpoint. 

\subsection{Model for monodisperse and bidisperse samples}
\label{t1mod}
As described in detail elsewhere~\cite{mcgraw12PRL, salez12EPJE, oron97RMP, craster09RMP}, the governing equation for the long time evolution of the profiles of monodisperse (\sti)\ samples is given by a single thin film equation (TFE):
\begin{equation}
\partial_th + \frac{\gamma}{3\eta}\partial_x\left(h^3\partial_x^3h\right) = 0\ ,
\label{hxt}
\end{equation}
where $h(x,t)$ is the distance between the substrate and the air-polymer interface at position $x$ and time $t$. This model for the evolution of the height profile is obtained by assuming a no slip boundary condition at the polymer-substrate interface, and no stress at the air-polymer interface. Additionally, it is assumed that the effects of inertia are negligible, that there is no viscoelastic effect, and that the height gradients are small. Numerically solving~\cite{salez12EPJE} a dimensionless form of Eq.~(\ref{hxt}) allows one to extract a capillary velocity, $\gamma/\eta$, by noting the single correspondence between the experimental coordinates $(x,t)$ and the dimensionless ones $(X,T)$:
\begin{equation}
u = \frac{x}{t^{1/4}} = \left(\frac{\gamma h_1^{\,3}}{3\eta}\right)^{1/4}\frac{X}{T^{1/4}}\ .
\label{xX}
\end{equation}
The assumptions leading to Eq.~(\ref{hxt}) were the subject of our previous investigations for monodisperse stepped films~\cite{mcgraw11SM, mcgraw12PRL, salez12EPJE, salez12PoF}, and their validity is now well established. For each of the height profiles shown in this work, we average more than 50 AFM scan lines, indicating that the error on an individual height measurement is given by the roughness of the sample ($\sim 0.3$\ nm RMS roughness as obtained from AFM measurements). Using this roughness as the dominant error term, we find that a typical error on the measurement of $\eta/\gamma$ is approximately 20\% of the measured value. Sample-to-sample variability can lead to a larger spread in measured values of $\eta/\gamma$, with the most important issue being the precision (here $\sim1\celc$) of the temperature control. 

We also describe homogeneous bidisperse stepped films (\stii) through TFE. When comparing theory to experiments, the only difference lies in the molecular weight model we use for each type of sample. For \sti\ samples, the molecular weight, $M$, is simply that of the PS used for sample preparation. However, if the polymer melt is not monodisperse the single parameter $M$ is not sufficient to characterize the melt. In this case, there is some distribution of chain lengths, and the viscosity will be a function of this distribution. As discussed in detail elsewhere~\cite{berryfox, graessley74APS, ferry, ninomiya63JPC, masuda70MAC, struglinski85MAC}, the viscosity of a polydisperse melt is known to follow the weight-averaged molecular weight. Thus, for \stii\  samples, we consider melts with weight fraction $\phi$ of molecular weight $M_l$ and with the remainder comprising chains of molecular weight $M_s$. The associated weight-averaged molecular weight is~\cite{rubincolby}:
\begin{equation}
M_{\textrm{[B]}} = M_l\phi + M_s(1-\phi)\ .
\label{bidMw}
\end{equation}

The mobility of polymer molecules is highly dependent on the chain length. On one hand, polymer chains that are much shorter than the segments of chain between entanglements, are in the so-called Rouse regime, where the viscosity scales linearly~\cite{rubincolby} with molecular weight. On the other hand, the reptation arguments due to de Gennes~\cite{degennesscaling} and Doi and Edwards~\cite{doied} account well for the high molecular weight dependence of the viscosity in linear chain systems where the chains are well entangled. Within the reptation model, a simple argument~\cite{rubincolby} can be used to predict that highly entangled monodisperse polymer melts have a viscosity that scales as $M^3$. In reality, there are other mechanisms~\cite{mcleish02AiP} that result in a stronger chain length dependence to the viscosity. It is commonly accepted~\cite{degennesscaling, rubincolby, berryfox, graessley74APS, colby87MAC} that monodisperse, high molecular weight polymer viscosity scales as $M^{3.4}$. Therefore, in both monodisperse and bidisperse cases, when the molecular weight is in the vicinity of a critical molecular weight for entanglement effects to become significant, $M_c$, the viscosity is described by the crossover function which smoothly connects the two molecular weight regimes~\cite{rubincolby}. Assuming that the surface tension of PS at 140\celc\ is constant\footnote{For the range of molecular weights used here, the surface tension changes by only a few percent~\cite{dee98AiP}. Given that the viscosity changes by six orders of magnitude over the same molecular weight range, the changes in capillary velocity, $\gamma/\eta$, are dominated by the changes in viscosity.} with~\cite{wu70JPC} $\gamma = 30$ mJ/m$^2$, one can write: 
\begin{equation}
\frac{\eta}{\gamma} = cM\left[\  1 + \left(\frac{M}{M_c}\right)^n\ \right]\ ,
\label{EGeq}
\end{equation}
where it is typically found~\cite{degennesscaling, rubincolby, berryfox, graessley74APS, colby87MAC} that $n = 2.4$ and, for PS, $M_c$ equals approximately~\cite{rubincolby, fetters3} 35\ kg/mol. 

\subsection{Model for spatially inhomogeneous samples}
\label{modA}
For spatially inhomogeneous samples (\stiii) we derive a thin film equation that is similar to TFE (see Eq.~(\ref{hxt})). The only difference is that now $M$ and thus $\eta$ are laterally inhomogeneous and depend on position and time through $h(x,t)$. However, we assume a perfect vertical mixing that leads to a local vertically weight-averaged molecular weight (see Eq.~(\ref{bidMw})):
\begin{equation}
M_{[\textrm{I}]}(h) =   \begin{cases}
M_1\frac{h_1}{h} + M_2\left(1-\frac{h_1}{h}\right) &\text{if } h \geq h^*\ ,  \\
M_1 &\text{if } h < h^*\ . \\
\end{cases}
\label{MwN}
\end{equation}
Note that $h^*$ is \textit{a priori} different from $h_1$. It is precisely defined through $h^*=h(x^*,t)$, such that $x^*=\max(u)$ and: 
\begin{equation}
\int_{-\infty}^{u}dx\ (h-h_1)=0\ .
\end{equation}
Thus, mass conservation of both molecular weights is ensured at all times:
\begin{equation}
\frac{\int_{-\infty}^{+\infty} dx\ M_{[\textrm{I}]}[h(x,t)]\ h(x,t)}{\int_{-\infty}^{+\infty} dx\ h(x,t)}=\frac{2M_1h_1+M_2h_2}{2h_1+h_2}\ .
\end{equation}
As in Eq.~(\ref{EGeq}), the viscosity is given by the crossover function which smoothly connects the Rouse and reptation regimes~\cite{rubincolby}:
\begin{equation}
\tilde{\eta}(h) =   \frac{\eta}{\eta_1} = \frac{\frac{M_{[\textrm{I}]}(h)}{M_c}+\left[\frac{M_{[\textrm{I}]}(h)}{M_c}\right]^{3.4}}{\frac{M_1}{M_c}+\left[\frac{M_1}{M_c}\right]^{3.4}}\ ,
\label{ettild}
\end{equation}
where the viscosity $\eta_1$ associated with the initial molecular weight $M_1$ of the lower film is chosen as a reference viscosity. Therefore, Eq.~(\ref{hxt}) is now replaced by:
\begin{equation}
\partial_th + \frac{\gamma}{3\eta_1}\partial_x\left[\frac{h^3}{\tilde{\eta}(h)}\partial_x^3h\right] = 0\ ,
\label{ITFE}
\end{equation}
which we will refer to as ITFE for `Inhomogeneous' TFE. Finally, nondimensionalizing Eq.~(\ref{ITFE}) through Eq.~(\ref{xX}), leads to:
\begin{equation}
\partial_TH + \partial_X\left[\frac{H^3}{\tilde{\eta}(H)}\partial_X^3H\right]= 0\ .
\label{ITFEnd}
\end{equation}
Using a finite-difference algorithm\cite{salez12EPJE}, one can compute the solution of Eq.~(\ref{ITFEnd}) for any initial aspect ratio and mass ratio of the step.

\section{Discussion}
\label{discuss}
This section is divided in three parts where we describe in turn: the capillary velocity of homogeneous samples; the observed self-similarity of inhomogeneous samples; and finally, the fitting and the capillary velocity of inhomogeneous samples.

\subsection{Capillary velocity of homogeneous samples}
\label{cvos}
In Fig.~\ref{typei} is shown the fit of a computed height profile  to that of an experimental profile from a monodisperse sample. The only fitting parameter is the horizontal scaling factor, which gives the capillary velocity, $\gamma/\eta$, according to Eq.~(\ref{xX}). 

Since the model described by Eq.~(\ref{hxt}) assumes only that the fluid is homogeneous, we expect that the predictions of TFE should also describe \stii\ samples. This numerical solution is thus fit to the data in Fig.~\ref{animals}(b) for the homogenous bidisperse \stii\ sample. The fit quality is excellent, thus giving the capillary velocity according to Eq.~(\ref{xX}). 
\begin{figure}[t!]
	\centering 
	\includegraphics{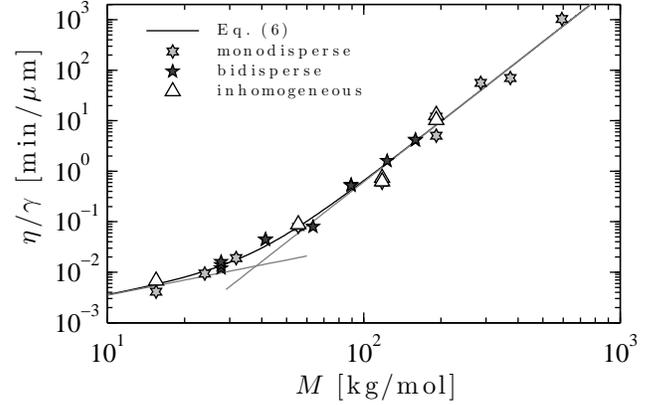}
	\caption{Inverse capillary velocities of all samples studied in this work, all annealed at 140\celc. For bidisperse (\stii) samples, we use $M = M_{\textrm{[B]}}$ according to Eq.~(\ref{bidMw}); for inhomogeneous (\stiii) samples, we use $M = M_1$, as described in Fig.~\ref{sampschem} and part~\ref{EGtN}. The line represents a fit to Eq.~(\ref{EGeq}) with $n = 3.0\pm0.6$ and $M_c = 38\pm12$\ kg/mol and $c = (3.5\pm2.6)\times10^{-4}$ min mol $\mu$m$^{-1}$ kg$^{-1}$; grey lines are the asymptotes of Eq.~(\ref{EGeq}). Each error represents a 90\% confidence interval assuming a 20\% error on each individual measurement of a capillary velocity (approximately the size of the data points), which is typical for these measurements. } 
	\label{EG} 
\end{figure} 

In Fig.~\ref{EG}, the $\eta/\gamma$ values for \sti\ and \stii\ samples are plotted as a function of the weight-averaged molecular weight. The values obtained for \stii\ samples are in agreement with the dependence observed for monodisperse samples. We therefore conclude that $M_{\textrm{[B]}}$ (see Eq.~(\ref{bidMw})) leads to a good predictor of the capillary velocity in bidisperse leveling stepped films. Moreover, we have experimentally demonstrated that the weight-averaged model for $M_\textrm{[I]}$, as defined by Eq.~(\ref{MwN}), is a reasonable approximation for the molecular weight characterization that should be used to determine the local viscosity in \stiii\ samples. We will focus on these inhomogeneous systems in the next part.

In Fig.~\ref{EG}, we show in addition to the $\eta/\gamma$ values measured, a fit to Eq.~(\ref{EGeq}), with $c$, $M_c$ and $n$ free fitting parameters. We find $n = 3.0\pm0.6$, $M_c = 38\pm12$\ kg/mol and $c = (3.5\pm2.6)\times10^{-4}$ min mol $\mu$m$^{-1}$ kg$^{-1}$, with errors expressing a 90\% confidence interval. This best fit value of $M_c$ is consistent with that obtained elsewhere, and the value of $c$ obtained is consistent with bulk expectations of the viscosity~\cite{bach2003} after multiplying by the surface tension of PS~\cite{wu70JPC} at 140\celc\ and adjusting for the temperature dependence of the viscosity through the Williams-Landel-Ferry equation~\cite{rubincolby, williams55JACS}.

The best fit value of $n$ we report is larger than the expected value, though we note that the deviation from the expected behaviour (\textit{i.e.} $n = 2.4$) is only significant for the highest molecular weight used (592 kg/mol). Furthermore, deviations from the expected exponent have been observed~\cite{ninomiya63JPC, colby87MAC} when the molecular weight is very large compared to $M_c$. For the purposes of this work, it is sufficient to note that bidisperse blends follow the same capillary velocity to molecular weight dependence as the monodisperse samples when the weight-averaged molecular weight is used. 

\subsection{Self-similarity of inhomogeneous samples}
\label{ssim}
\begin{figure}[t!]
	\centering 
	\includegraphics{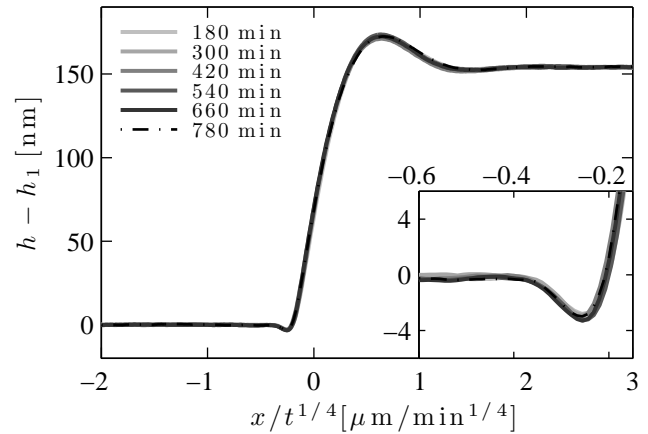}
	\caption{Experimental data from a \stiii\ sample with $\{M_1, M_2\} = \{192, 55.5\}$\ kg/mol and $\{h_1, h_2\} = \{66, 153\}$\ nm annealed at 140\celc\ for times indicated. We plot the measured height as a function of the self-similar variable $xt^{-1/4}$.} 
	\label{selfsim} 
\end{figure}

Having shown in Figs.~\ref{animals} and~\ref{bdplot} and in part~\ref{gogc} that \stiii\ sample profiles evolve qualitatively as we have expected based on their mobility distribution, it is pertinent to ask whether the observed behaviours are transient or not. To test this question, we have annealed the \stiii\ samples in Fig.~\ref{animals} for various times. We have also prepared a second sample of \stiii\ with $\{M_1, M_2\} = \{192, 55.5\}$\ kg/mol, but with geometry $\{h_1, h_2\} = \{66, 153\}$\ nm. In Fig.~\ref{selfsim}, we show the temporal evolution of this height profile in scaled units, demonstrating the self-similarity of the evolution. The inset shows in detail the dip region. Consistent with the discussion in part~\ref{gogc}, this sample shows an enhanced \RR\ value -- the thick side of the sample has transported more material as compared to a homogeneous sample with the same geometry. We stress that this different geometry also serves to validate the molecular weight model of Eq.~(\ref{MwN}), and that the fit quality using ITFE (see Eq.~(\ref{ITFEnd})) is comparable with those shown in Figs.~\ref{animals}(a) and~\ref{animals}(b).

In addition to the data shown in Figs.~\ref{animals}(a) and~\ref{animals}(c) for an annealing time of $t = 240$\ min, the height profiles were also obtained for $t = 120\textrm{ and }360$\ min of annealing at $140\celc$ for those samples (not shown). Similarly, the height profile of the sample shown in Fig.~\ref{animals}(b) was also obtained at $t = 120$\ min. While the profiles are shown only at a single time for clarity, we note that they are \emph{all} self-similar in $xt^{-1/4}$ at the times shown. 

The fact that all the \stiii\ profiles we have measured satisfy the self-similarity condition provides strong evidence that a single driving force dominates the evolution of our samples for the time scales considered. Of specific relevance to the mixed $M$ samples schematically shown in Fig.~\ref{sampschem}, there are two ways in which entropic mixing might lead to a breakdown of the observed self-similarity. 

First, in the as-prepared state, the molecular weights, $\{M_1, M_2\}$, on the thick side of the stepped film are segregated, see Figs.~\ref{sampschem}(a) and~\ref{sampschem}(b.i). As such, the vertical viscosity profile continuously varies as the molecules from the component films mix~\cite{kramer84POL, greenMAC86, liuMAC93}. Therefore, the height profile cannot be self-similar in $xt^{-1/4}$ at early times since the viscosity profile depends explicitly on time, see Fig.~\ref{sampschem}(b.i) -~\ref{sampschem}(b.iv). We can make an estimate for the time scales over which this transient vertical mixing regime might last by making a simple reptation argument. We use: the reptation time~\cite{bach2003} for 192 kg/mol PS at 140\celc; the assumption that the polymer coil size, $R_g$, is roughly 10 times smaller than the typical film heights used here~\cite{cotton74MAC}; a compositionally dependent reptation time, $\tau_\phi$, using the theory developed by Kramer and co-workers~\cite{kramer84POL} (Eq. (23) in their work, with $M_1/M_2 \approx 4$ for the PS used in Fig.~\ref{animals}, and the average concentration, $\langle\phi\rangle = 1/2$); and the assumption that our samples are well mixed after a chain diffuses over distances comparable to typical film thicknesses: $t_\textrm{mix} = (h/R_g)^2\tau_\phi$. With these assumptions, we find that it should take $t_\textrm{mix} \approx 80$\ min to reach a vertically well mixed state for the $\{55.5, 192\}$ and $\{192, 55.5\}$\ kg/mol \stiii\ samples in Fig.~\ref{animals}(a) and~\ref{animals}(c). This time corresponds well with the fact that we observe self-similar profiles at $t = 240$\ min for these samples. 

Secondly, lateral diffusion can lead to mixing. However, this diffusion is much slower than the hydrodynamic processes we observe here. Making the same argument as in the previous paragraph, we expect that it would take $10^5$\ min to mix the two molecular weights over the typical flow distances we observe in this experiment ($\sim 5\ \mu$m), we can thus ignore lateral diffusion on the time scales considered here.

The observation of self-similarity in $xt^{-1/4}$ gives support for the idea that the samples are vertically well mixed; therefore, the Poiseuille flow schematically shown in Fig.~\ref{sampschem}(b.iv) and used in the derivation of Eq.~(\ref{ITFEnd}) is a good assumption. Furthermore, the observed self-similarity suggests that lateral mixing of the two different molecular weight distributions is not diffusively driven, but is instead predominantly accomplished through the hydrodynamic capillary flow. 

\subsection{Comparison with model and capillary velocity for inhomogeneous samples}
\label{EGtN} 
As described in part~\ref{modA}, the profiles computed according to Eq.~(\ref{ITFEnd}) are dependent on $\{M_1, M_2\}$ (see Fig.~\ref{sampschem}) and on the dimensionless heights. Having measured the experimental $\{h_1, h_2\}$, and knowing the molecular weights used in the samples, we see that the relative viscosity may be determined through the vertically weight-averaged molecular weight of a bidisperse blend (see Eq.~(\ref{ettild})). Therefore, the temporal evolution of ITFE (see Eq.~(\ref{ITFEnd})) may be computed uniquely for a given sample. 

Validation for the ITFE model (see Eqs.~(\ref{MwN}), (\ref{ettild}), (\ref{ITFE}), and (\ref{ITFEnd})) is found in the fact that it better predicts the measured height profiles of \stiii\ samples than the TFE model. This is demonstrated in Figs.~\ref{animals}(a) and~\ref{animals}(c) where we show a comparison between numerical solutions of TFE (see Eq.~(\ref{hxt})) and ITFE (see Eq.~(\ref{ITFE}))  which have been fit to the experimental data. We note that fits to ITFE for samples with $M_1>M_2$ (Fig.~\ref{animals}(c)) are better than for the case where $M_1<M_2$ (Fig.~\ref{animals}(a)). 
For the profiles in Fig.~\ref{animals}(a), there is a small but systematic deviation near the bump. This type of deviation is also visible in similar samples with a different $\{M_1, M_2\}$, when $M_1<M_2$. We speculate that this deviation may result from a small vertical inhomogeneity, which invalidates the assumption of Poiseuille flow, though the details remain elusive. Notwithstanding this negligible deviation near the bump for \stiii\ samples, we find that the agreement between theory and experiments is consistently better with the use of ITFE (see Eq.~(\ref{ITFE})) than with TFE (see Eq.~(\ref{hxt})). The ITFE model explicitly takes into account the laterally inhomogeneous mobility of the samples, which is not done by the TFE model. 

A horizontal stretch factor is the only free parameter needed for comparison between height profiles measured from \sti\ and \stii\ samples and corresponding solutions of TFE (see  Eq.~(\ref{hxt})). Similarly, the only parameter needed for comparison between the height profiles measured from \stiii\ samples, and the appropriate solutions of ITFE (see Eq.~(\ref{ITFEnd})), is a horizontal stretching parameter. In analogy with Eq.~(\ref{xX}), the ITFE model leads to:
\begin{equation}
\frac{x}{t^{1/4}} = \left(\frac{\gamma h_1^{\,3}}{3\eta_1}\right)^{1/4}\frac{X}{T^{1/4}}\ .
\label{xXITFE}
\end{equation}
We stress that this correspondence would not be possible if the experimentally observed profiles were not self-similar. 

Thus, a fit to the ITFE profiles we compute (see Eq.~(\ref{xXITFE})) leads to a measure of the capillary velocity, $\gamma/\eta_1$, of the unmixed thin side of our spatially inhomogeneous, \stiii\ stepped films. For the profile shown in Fig.~\ref{animals}(a), we extract the capillary velocity of 55.5\ kg/mol PS, while corresponding fits to the profiles in Figs.~\ref{animals}(c) and~\ref{selfsim} give the capillary velocity of 192\ kg/mol PS. We have additionally prepared analogous \stiii\ samples to the ones shown in Figs.~\ref{animals}(a) and~\ref{animals}(c), but instead using 15.5 and 118\ kg/mol PS. These profiles show the same qualitative features as described in parts~\ref{gogc} and \ref{ssim}, and are included in Fig.~\ref{bdplot}. In Fig.~\ref{EG}, we plot with white triangles the values of $\eta_1/\gamma$ extracted from fits to all \stiii\ samples studied here. 

The molecular weight dependence of of the capillary velocity extracted from \stiii\ samples is consistent with the dependence observed in \sti\ samples, for which the model used is well established~\cite{mcgraw11SM, mcgraw12PRL, salez12EPJE, salez12PoF}. Furthermore, the extracted capillary velocity is not dependent on the sample geometry. The independence of capillary velocity on the geometry is demonstrated by the same value of $\eta/\gamma$ obtained for \stiii\ samples of identical $\{M_1, M_2\}$, but different $\{h_1, h_2\}$ (see Figs.~\ref{animals}(c) and~\ref{selfsim}). Since the dependence of capillary velocity in \stiii\ samples on molecular weight is consistent with that observed in \sti\ samples, it is thus also consistent with homogeneous bidisperse (\stii) samples as discussed in part~\ref{cvos}. Therefore, the ITFE model for the self-similar evolution of \stiii\ samples is validated as well. 

\section{Conclusion}
In this study, we have examined the effects of chain length distribution on the self-similar evolution of nanoscopic polymer stepped films. In the first case, we prepared stepped films from bidisperse polymer solutions such that the resulting films were homogeneous while containing chains of two differing lengths. We then observed a self-similar leveling toward the equilibrium flat film, as in the previous studies on homogeneous monodisperse samples~\cite{mcgraw11SM, mcgraw12PRL, salez12EPJE, salez12PoF}. Fitting the experimental data to computational profiles allowed us to measure the capillary velocity, $\gamma/\eta$, of bidisperse polymer blends. When plotted as a function of the weight-averaged molecular weight of the blend, we found that the molecular weight dependence of $\eta/\gamma$ in bidisperse blends is consistent with the observed dependence of monodisperse samples, thus demonstrating the validity of this averaging law in the case of polymer nanofilms. To our knowledge, this is the first time that the relation between molecular weight and viscosity has been tested for a nanoscopic, bidisperse system. 

Then, stepped films were prepared such that, in the initial state, two molecular weights were spatially well segregated. This segregation was accomplished by stacking two films of different molecular weights atop one another. After an initial transient period, which was not studied here, we found that the evolution of the stepped films toward equilibrium is self-similar. This finding may be surprising given that there is an entropic drive to laterally homogenize the films. However, in the explored temporal range, the lateral diffusive process is slow in comparison to the driven flow resulting from capillary forces. The lubrication model developed is based on a vertically well mixed bidisperse blend, which allows us to assume a local Poiseuille flow. The numerical calculations from this laterally inhomogeneous mobility model are in good agreement with the experimentally measured profiles, and lead to a measure of the capillary velocity of an unmixed portion of the stepped films. The molecular weight dependence of $\eta/\gamma$ on molecular weight extracted from these inhomogeneous samples is consistent with the observed homogeneous dependence.  Having deliberately prepared samples with mobility gradients, we hope that the results and models presented in this study may guide the interpretation of measurements made on confined and glassy polymer stepped films. 

\section{Acknowledgements}
The authors thank NSERC of Canada, the \'Ecole Normale Sup\'erieure of Paris, the Fondation Langlois, the German Research Foundation (DFG) under Grant No. BA3406/2, the Chaire Total-ESPCI ParisTech, and the Saint Gobain Fellowship for financial support. The authors also thank Matilda Backholm, Michael Benzaquen, and Paul Fowler for technical assistance and interesting discussions. 
\footnotesize{

\begin{mcitethebibliography}{57}
\providecommand*{\natexlab}[1]{#1}
\providecommand*{\mciteSetBstSublistMode}[1]{}
\providecommand*{\mciteSetBstMaxWidthForm}[2]{}
\providecommand*{\mciteBstWouldAddEndPuncttrue}
  {\def\EndOfBibitem{\unskip.}}
\providecommand*{\mciteBstWouldAddEndPunctfalse}
  {\let\EndOfBibitem\relax}
\providecommand*{\mciteSetBstMidEndSepPunct}[3]{}
\providecommand*{\mciteSetBstSublistLabelBeginEnd}[3]{}
\providecommand*{\EndOfBibitem}{}
\mciteSetBstSublistMode{f}
\mciteSetBstMaxWidthForm{subitem}
{(\emph{\alph{mcitesubitemcount}})}
\mciteSetBstSublistLabelBeginEnd{\mcitemaxwidthsubitemform\space}
{\relax}{\relax}

\bibitem[Tsui and Russell(2008)]{tsui08TXT}
\emph{Polymer Thin Films}, ed. O.~Tsui and T.~Russell, World Scientific,
  2008\relax
\mciteBstWouldAddEndPuncttrue
\mciteSetBstMidEndSepPunct{\mcitedefaultmidpunct}
{\mcitedefaultendpunct}{\mcitedefaultseppunct}\relax
\EndOfBibitem
\bibitem[Jaeger(2002)]{jaeger02TXT}
R.~Jaeger, \emph{Introduction to microelectronic fabrication}, Prentice Hall,
  2002\relax
\mciteBstWouldAddEndPuncttrue
\mciteSetBstMidEndSepPunct{\mcitedefaultmidpunct}
{\mcitedefaultendpunct}{\mcitedefaultseppunct}\relax
\EndOfBibitem
\bibitem[Tang \emph{et~al.}(2008)Tang, Lennon, Fredrickson, Kramer, and
  Hawker]{tang08SCI}
C.~Tang, E.~Lennon, G.~Fredrickson, E.~Kramer and C.~Hawker, \emph{Science},
  2008, \textbf{322}, 429\relax
\mciteBstWouldAddEndPuncttrue
\mciteSetBstMidEndSepPunct{\mcitedefaultmidpunct}
{\mcitedefaultendpunct}{\mcitedefaultseppunct}\relax
\EndOfBibitem
\bibitem[Park \emph{et~al.}(2009)Park, Lee, Xu, Kim, Hong, Jeong, Xu, and
  Russell]{park09SCI}
S.~Park, D.~Lee, J.~Xu, B.~Kim, S.~Hong, U.~Jeong, T.~Xu and T.~Russell,
  \emph{Science}, 2009, \textbf{323}, 1030\relax
\mciteBstWouldAddEndPuncttrue
\mciteSetBstMidEndSepPunct{\mcitedefaultmidpunct}
{\mcitedefaultendpunct}{\mcitedefaultseppunct}\relax
\EndOfBibitem
\bibitem[Craster and Matar(2009)]{craster09RMP}
M.~Craster and O.~Matar, \emph{Reviews of Modern Physics}, 2009, \textbf{81},
  1131\relax
\mciteBstWouldAddEndPuncttrue
\mciteSetBstMidEndSepPunct{\mcitedefaultmidpunct}
{\mcitedefaultendpunct}{\mcitedefaultseppunct}\relax
\EndOfBibitem
\bibitem[Eijkel and van~den Berg(2005)]{eijkel05MNF}
J.~Eijkel and A.~van~den Berg, \emph{Microfluid Nanofluid}, 2005, \textbf{1},
  249\relax
\mciteBstWouldAddEndPuncttrue
\mciteSetBstMidEndSepPunct{\mcitedefaultmidpunct}
{\mcitedefaultendpunct}{\mcitedefaultseppunct}\relax
\EndOfBibitem
\bibitem[{d}e Gennes(1979)]{degennesscaling}
P.~{d}e Gennes, \emph{Scaling Concepts in Polymer Physics}, Cornell University
  Press, 1979\relax
\mciteBstWouldAddEndPuncttrue
\mciteSetBstMidEndSepPunct{\mcitedefaultmidpunct}
{\mcitedefaultendpunct}{\mcitedefaultseppunct}\relax
\EndOfBibitem
\bibitem[Matsen(2002)]{matsenJPCM02}
M.~Matsen, \emph{Journal of Physics: Condensed Matter}, 2002, \textbf{14},
  R21\relax
\mciteBstWouldAddEndPuncttrue
\mciteSetBstMidEndSepPunct{\mcitedefaultmidpunct}
{\mcitedefaultendpunct}{\mcitedefaultseppunct}\relax
\EndOfBibitem
\bibitem[Fakhraai and Forrest(2008)]{fakhraai08SCI}
Z.~Fakhraai and J.~Forrest, \emph{Science}, 2008, \textbf{319}, 600\relax
\mciteBstWouldAddEndPuncttrue
\mciteSetBstMidEndSepPunct{\mcitedefaultmidpunct}
{\mcitedefaultendpunct}{\mcitedefaultseppunct}\relax
\EndOfBibitem
\bibitem[Paeng \emph{et~al.}(2011)Paeng, Swallen, and Ediger]{paeng11JACS}
K.~Paeng, S.~Swallen and M.~Ediger, \emph{Journal of the American Chemical
  Society}, 2011, \textbf{133}, 8444\relax
\mciteBstWouldAddEndPuncttrue
\mciteSetBstMidEndSepPunct{\mcitedefaultmidpunct}
{\mcitedefaultendpunct}{\mcitedefaultseppunct}\relax
\EndOfBibitem
\bibitem[Keddie \emph{et~al.}(1994)Keddie, Jones, and Cory]{keddie94EPL}
J.~Keddie, R.~Jones and R.~Cory, \emph{Europhysics Letters}, 1994, \textbf{27},
  59\relax
\mciteBstWouldAddEndPuncttrue
\mciteSetBstMidEndSepPunct{\mcitedefaultmidpunct}
{\mcitedefaultendpunct}{\mcitedefaultseppunct}\relax
\EndOfBibitem
\bibitem[Forrest and Dalnoki-Veress(2001)]{forrest01ACIS}
J.~Forrest and K.~Dalnoki-Veress, \emph{Advances in Colloid and Interface
  Science}, 2001, \textbf{94}, 167\relax
\mciteBstWouldAddEndPuncttrue
\mciteSetBstMidEndSepPunct{\mcitedefaultmidpunct}
{\mcitedefaultendpunct}{\mcitedefaultseppunct}\relax
\EndOfBibitem
\bibitem[Seemann \emph{et~al.}(2006)Seemann, Jacobs, Landfester, and
  Herminghaus]{seemann06JPS}
R.~Seemann, K.~Jacobs, K.~Landfester and S.~Herminghaus, \emph{Journal of
  Polymer Science Part B: Polymer Physics}, 2006, \textbf{44}, 2968\relax
\mciteBstWouldAddEndPuncttrue
\mciteSetBstMidEndSepPunct{\mcitedefaultmidpunct}
{\mcitedefaultendpunct}{\mcitedefaultseppunct}\relax
\EndOfBibitem
\bibitem[Ellison and Torkelson(2003)]{ellisonNATM03}
C.~Ellison and J.~Torkelson, \emph{Nature Materials}, 2003, \textbf{2},
  695\relax
\mciteBstWouldAddEndPuncttrue
\mciteSetBstMidEndSepPunct{\mcitedefaultmidpunct}
{\mcitedefaultendpunct}{\mcitedefaultseppunct}\relax
\EndOfBibitem
\bibitem[Si \emph{et~al.}(2005)Si, Massa, Dalnoki-Veress, Brown, and
  Jones]{allan}
L.~Si, M.~Massa, K.~Dalnoki-Veress, H.~Brown and R.~Jones, \emph{Phys. Rev.
  Lett.}, 2005, \textbf{94}, 127801\relax
\mciteBstWouldAddEndPuncttrue
\mciteSetBstMidEndSepPunct{\mcitedefaultmidpunct}
{\mcitedefaultendpunct}{\mcitedefaultseppunct}\relax
\EndOfBibitem
\bibitem[O'{C}onnell and McKenna(2006)]{oconnellSCI2006}
P.~O'{C}onnell and G.~McKenna, \emph{Science}, 2006, \textbf{307}, 1760\relax
\mciteBstWouldAddEndPuncttrue
\mciteSetBstMidEndSepPunct{\mcitedefaultmidpunct}
{\mcitedefaultendpunct}{\mcitedefaultseppunct}\relax
\EndOfBibitem
\bibitem[Bodiguel and Fretigny(2006)]{Bodiguel2006}
H.~Bodiguel and C.~Fretigny, \emph{Physical Review Letters}, 2006, \textbf{97},
  266105\relax
\mciteBstWouldAddEndPuncttrue
\mciteSetBstMidEndSepPunct{\mcitedefaultmidpunct}
{\mcitedefaultendpunct}{\mcitedefaultseppunct}\relax
\EndOfBibitem
\bibitem[Shin \emph{et~al.}(2007)Shin, Obukhov, Chen, Huh, Hwang, Mok,
  Dobriyal, Thiyagarajan, and Russell]{shinNATM07}
K.~Shin, S.~Obukhov, J.-T. Chen, J.~Huh, Y.~Hwang, S.~Mok, P.~Dobriyal,
  P.~Thiyagarajan and T.~Russell, \emph{Nature Materials}, 2007, \textbf{6},
  961\relax
\mciteBstWouldAddEndPuncttrue
\mciteSetBstMidEndSepPunct{\mcitedefaultmidpunct}
{\mcitedefaultendpunct}{\mcitedefaultseppunct}\relax
\EndOfBibitem
\bibitem[Israelachvili(2011)]{isf11TXT}
J.~Israelachvili, \emph{Intermolecular and Surface Forces}, Academic Press, 3rd
  edn, 2011\relax
\mciteBstWouldAddEndPuncttrue
\mciteSetBstMidEndSepPunct{\mcitedefaultmidpunct}
{\mcitedefaultendpunct}{\mcitedefaultseppunct}\relax
\EndOfBibitem
\bibitem[Reiter(1992)]{reiter92PRL}
G.~Reiter, \emph{Physical Review Letters}, 1992, \textbf{68}, 75\relax
\mciteBstWouldAddEndPuncttrue
\mciteSetBstMidEndSepPunct{\mcitedefaultmidpunct}
{\mcitedefaultendpunct}{\mcitedefaultseppunct}\relax
\EndOfBibitem
\bibitem[Reiter(2008)]{reiter08TXT}
G.~Reiter, in \emph{Soft Matter Characterization}, ed. R.~Borsali and
  R.~Pecora, Springer-Verlag, 2008, ch. 25: Visualizing Properties of Polymers
  at Interfaces\relax
\mciteBstWouldAddEndPuncttrue
\mciteSetBstMidEndSepPunct{\mcitedefaultmidpunct}
{\mcitedefaultendpunct}{\mcitedefaultseppunct}\relax
\EndOfBibitem
\bibitem[Seemann \emph{et~al.}(2001)Seemann, Herminghaus, and
  Jacobs]{seemann01PRL}
R.~Seemann, S.~Herminghaus and K.~Jacobs, \emph{Physical Review Letters}, 2001,
  \textbf{87}, 196101\relax
\mciteBstWouldAddEndPuncttrue
\mciteSetBstMidEndSepPunct{\mcitedefaultmidpunct}
{\mcitedefaultendpunct}{\mcitedefaultseppunct}\relax
\EndOfBibitem
\bibitem[Seemann \emph{et~al.}(2001)Seemann, Herminhaus, and
  Jacobs]{seeman01PRLb}
R.~Seemann, S.~Herminhaus and K.~Jacobs, \emph{Physical Review Letters}, 2001,
  \textbf{86}, 5534\relax
\mciteBstWouldAddEndPuncttrue
\mciteSetBstMidEndSepPunct{\mcitedefaultmidpunct}
{\mcitedefaultendpunct}{\mcitedefaultseppunct}\relax
\EndOfBibitem
\bibitem[Dalnoki-Veress \emph{et~al.}(2001)Dalnoki-Veress, Forrest, Murray,
  Gigault, and Dutcher]{dalnoki01PRE}
K.~Dalnoki-Veress, J.~Forrest, C.~Murray, C.~Gigault and J.~Dutcher,
  \emph{Physical Review E}, 2001, \textbf{63}, 031801\relax
\mciteBstWouldAddEndPuncttrue
\mciteSetBstMidEndSepPunct{\mcitedefaultmidpunct}
{\mcitedefaultendpunct}{\mcitedefaultseppunct}\relax
\EndOfBibitem
\bibitem[Roth and Dutcher(2006)]{roth06JPSB}
C.~Roth and J.~Dutcher, \emph{Journal of Polymer Science Part B: Polymer
  Physics}, 2006, \textbf{44}, 3011\relax
\mciteBstWouldAddEndPuncttrue
\mciteSetBstMidEndSepPunct{\mcitedefaultmidpunct}
{\mcitedefaultendpunct}{\mcitedefaultseppunct}\relax
\EndOfBibitem
\bibitem[Barbero and Steiner(2009)]{barbero2009}
D.~Barbero and U.~Steiner, \emph{Physical Review Letters}, 2009, \textbf{102},
  248303\relax
\mciteBstWouldAddEndPuncttrue
\mciteSetBstMidEndSepPunct{\mcitedefaultmidpunct}
{\mcitedefaultendpunct}{\mcitedefaultseppunct}\relax
\EndOfBibitem
\bibitem[Thomas \emph{et~al.}(2011)Thomas, Chenneviere, Reiter, and
  Steiner]{thomas11PRE}
K.~Thomas, A.~Chenneviere, G.~Reiter and U.~Steiner, \emph{Physical Review E},
  2011, \textbf{83}, 1\relax
\mciteBstWouldAddEndPuncttrue
\mciteSetBstMidEndSepPunct{\mcitedefaultmidpunct}
{\mcitedefaultendpunct}{\mcitedefaultseppunct}\relax
\EndOfBibitem
\bibitem[Closa \emph{et~al.}(2011)Closa, Ziebert, and Rapha\"el]{Closa2011}
F.~Closa, F.~Ziebert and E.~Rapha\"el, \emph{Phys. Rev. E}, 2011,
  \textbf{{\bf83}}, 051603\relax
\mciteBstWouldAddEndPuncttrue
\mciteSetBstMidEndSepPunct{\mcitedefaultmidpunct}
{\mcitedefaultendpunct}{\mcitedefaultseppunct}\relax
\EndOfBibitem
\bibitem[McGraw \emph{et~al.}(2011)McGraw, Jago, and
  Dalnoki-Veress]{mcgraw11SM}
J.~McGraw, N.~Jago and K.~Dalnoki-Veress, \emph{Soft Matter}, 2011, \textbf{7},
  7832\relax
\mciteBstWouldAddEndPuncttrue
\mciteSetBstMidEndSepPunct{\mcitedefaultmidpunct}
{\mcitedefaultendpunct}{\mcitedefaultseppunct}\relax
\EndOfBibitem
\bibitem[McGraw \emph{et~al.}(2012)McGraw, Salez, B\"aumchen, Rapha\"el, and
  Dalnoki-Veress]{mcgraw12PRL}
J.~McGraw, T.~Salez, O.~B\"aumchen, E.~Rapha\"el and K.~Dalnoki-Veress,
  \emph{Physical Review Letters}, 2012, \textbf{109}, 128303\relax
\mciteBstWouldAddEndPuncttrue
\mciteSetBstMidEndSepPunct{\mcitedefaultmidpunct}
{\mcitedefaultendpunct}{\mcitedefaultseppunct}\relax
\EndOfBibitem
\bibitem[Salez \emph{et~al.}(2012)Salez, McGraw, Cormier, B\"aumchen,
  Dalnoki-Veress, and Rapha\"el]{salez12EPJE}
T.~Salez, J.~McGraw, S.~Cormier, O.~B\"aumchen, K.~Dalnoki-Veress and
  E.~Rapha\"el, \emph{European Physical Journal E}, 2012, \textbf{35},
  114\relax
\mciteBstWouldAddEndPuncttrue
\mciteSetBstMidEndSepPunct{\mcitedefaultmidpunct}
{\mcitedefaultendpunct}{\mcitedefaultseppunct}\relax
\EndOfBibitem
\bibitem[Salez \emph{et~al.}(2012)Salez, McGraw, B\"aumchen, Dalnoki-Veress,
  and Rapha\"el]{salez12PoF}
T.~Salez, J.~McGraw, O.~B\"aumchen, K.~Dalnoki-Veress and E.~Rapha\"el,
  \emph{Physics of Fluids}, 2012, \textbf{24}, 102111\relax
\mciteBstWouldAddEndPuncttrue
\mciteSetBstMidEndSepPunct{\mcitedefaultmidpunct}
{\mcitedefaultendpunct}{\mcitedefaultseppunct}\relax
\EndOfBibitem
\bibitem[Oron \emph{et~al.}(1997)Oron, Davis, and Bankoff]{oron97RMP}
A.~Oron, S.~Davis and S.~Bankoff, \emph{Reviews of Modern Physics}, 1997,
  \textbf{69}, 931\relax
\mciteBstWouldAddEndPuncttrue
\mciteSetBstMidEndSepPunct{\mcitedefaultmidpunct}
{\mcitedefaultendpunct}{\mcitedefaultseppunct}\relax
\EndOfBibitem
\bibitem[Struglinski and Graessley(1985)]{struglinski85MAC}
M.~Struglinski and W.~Graessley, \emph{Macromolecules}, 1985, \textbf{18},
  2630\relax
\mciteBstWouldAddEndPuncttrue
\mciteSetBstMidEndSepPunct{\mcitedefaultmidpunct}
{\mcitedefaultendpunct}{\mcitedefaultseppunct}\relax
\EndOfBibitem
\bibitem[Kramer \emph{et~al.}(1984)Kramer, Green, and Palmstrom]{kramer84POL}
E.~Kramer, P.~Green and C.~Palmstrom, \emph{Polymer}, 1984, \textbf{25},
  473\relax
\mciteBstWouldAddEndPuncttrue
\mciteSetBstMidEndSepPunct{\mcitedefaultmidpunct}
{\mcitedefaultendpunct}{\mcitedefaultseppunct}\relax
\EndOfBibitem
\bibitem[Green and Kramer(1986)]{greenMAC86}
P.~Green and E.~Kramer, \emph{Macromolecules}, 1986, \textbf{19}, 1108\relax
\mciteBstWouldAddEndPuncttrue
\mciteSetBstMidEndSepPunct{\mcitedefaultmidpunct}
{\mcitedefaultendpunct}{\mcitedefaultseppunct}\relax
\EndOfBibitem
\bibitem[Liu \emph{et~al.}(1993)Liu, Reiter, Kunz, and Stamm]{liuMAC93}
Y.~Liu, G.~Reiter, K.~Kunz and M.~Stamm, \emph{Macromolecules}, 1993,
  \textbf{26}, 2134\relax
\mciteBstWouldAddEndPuncttrue
\mciteSetBstMidEndSepPunct{\mcitedefaultmidpunct}
{\mcitedefaultendpunct}{\mcitedefaultseppunct}\relax
\EndOfBibitem
\bibitem[Karim \emph{et~al.}(1994)Karim, Felcher, and Russell]{karimMAC94}
A.~Karim, G.~Felcher and T.~Russell, \emph{Macromolecules}, 1994, \textbf{27},
  6973\relax
\mciteBstWouldAddEndPuncttrue
\mciteSetBstMidEndSepPunct{\mcitedefaultmidpunct}
{\mcitedefaultendpunct}{\mcitedefaultseppunct}\relax
\EndOfBibitem
\bibitem[Pierce \emph{et~al.}(2011)Pierce, Perahia, and Grest]{pierce11EPL}
F.~Pierce, D.~Perahia and G.~Grest, \emph{Europhysics Letters}, 2011,
  \textbf{95}, 46001\relax
\mciteBstWouldAddEndPuncttrue
\mciteSetBstMidEndSepPunct{\mcitedefaultmidpunct}
{\mcitedefaultendpunct}{\mcitedefaultseppunct}\relax
\EndOfBibitem
\bibitem[Wang \emph{et~al.}(2011)Wang, Cheng, and Wang]{wang11JoR}
Y.~Wang, S.~Cheng and S.-Q. Wang, \emph{Journal of Rheology}, 2011,
  \textbf{55}, 1247\relax
\mciteBstWouldAddEndPuncttrue
\mciteSetBstMidEndSepPunct{\mcitedefaultmidpunct}
{\mcitedefaultendpunct}{\mcitedefaultseppunct}\relax
\EndOfBibitem
\bibitem[Cotton \emph{et~al.}(1974)Cotton, Decker, Benoit, Farnoux, Higgins,
  Jannink, Ober, Picot, and des Cloizeaux]{cotton74MAC}
J.~Cotton, D.~Decker, H.~Benoit, B.~Farnoux, J.~Higgins, G.~Jannink, R.~Ober,
  C.~Picot and J.~des Cloizeaux, \emph{Macromolecules}, 1974, \textbf{7},
  863\relax
\mciteBstWouldAddEndPuncttrue
\mciteSetBstMidEndSepPunct{\mcitedefaultmidpunct}
{\mcitedefaultendpunct}{\mcitedefaultseppunct}\relax
\EndOfBibitem
\bibitem[Kramer and Berger(1990)]{kramercrrev2}
E.~Kramer and L.~Berger, \emph{Advances in Polymer Science}, 1990,
  \textbf{91/92}, 1--68\relax
\mciteBstWouldAddEndPuncttrue
\mciteSetBstMidEndSepPunct{\mcitedefaultmidpunct}
{\mcitedefaultendpunct}{\mcitedefaultseppunct}\relax
\EndOfBibitem
\bibitem[{d}e Gennes \emph{et~al.}(2003){d}e Gennes, Brochard-Wyart, and
  Qu\'er\'e]{degennes03TXT}
P.~{d}e Gennes, F.~Brochard-Wyart and D.~Qu\'er\'e, \emph{Capillarity and
  Wetting Phenomena: Drops, Bubbles, Pearls, Waves}, Springer, New York,
  2003\relax
\mciteBstWouldAddEndPuncttrue
\mciteSetBstMidEndSepPunct{\mcitedefaultmidpunct}
{\mcitedefaultendpunct}{\mcitedefaultseppunct}\relax
\EndOfBibitem
\bibitem[Berry and Fox(1968)]{berryfox}
G.~Berry and T.~Fox, \emph{Advances in Polymer Science}, 1968, \textbf{5},
  261\relax
\mciteBstWouldAddEndPuncttrue
\mciteSetBstMidEndSepPunct{\mcitedefaultmidpunct}
{\mcitedefaultendpunct}{\mcitedefaultseppunct}\relax
\EndOfBibitem
\bibitem[Graessley(1974)]{graessley74APS}
W.~Graessley, \emph{Advances in Polymer Science}, 1974, \textbf{16}, 1\relax
\mciteBstWouldAddEndPuncttrue
\mciteSetBstMidEndSepPunct{\mcitedefaultmidpunct}
{\mcitedefaultendpunct}{\mcitedefaultseppunct}\relax
\EndOfBibitem
\bibitem[Ferry(1980)]{ferry}
J.~Ferry, \emph{Viscoelastic {P}roperties of {P}olymers, 3rd ed.}, John Wiley
  \& Sons, Inc., 1980\relax
\mciteBstWouldAddEndPuncttrue
\mciteSetBstMidEndSepPunct{\mcitedefaultmidpunct}
{\mcitedefaultendpunct}{\mcitedefaultseppunct}\relax
\EndOfBibitem
\bibitem[Ninomiya \emph{et~al.}(1963)Ninomiya, Ferry, and
  Oyanagi]{ninomiya63JPC}
K.~Ninomiya, J.~Ferry and Y.~Oyanagi, \emph{Journal of Physical Chemistry},
  1963, \textbf{67}, 2297\relax
\mciteBstWouldAddEndPuncttrue
\mciteSetBstMidEndSepPunct{\mcitedefaultmidpunct}
{\mcitedefaultendpunct}{\mcitedefaultseppunct}\relax
\EndOfBibitem
\bibitem[Masuda \emph{et~al.}(1970)Masuda, Kitagawa, Inoue, and
  Onogi]{masuda70MAC}
T.~Masuda, K.~Kitagawa, T.~Inoue and S.~Onogi, \emph{Macromolecules}, 1970,
  \textbf{3}, 116\relax
\mciteBstWouldAddEndPuncttrue
\mciteSetBstMidEndSepPunct{\mcitedefaultmidpunct}
{\mcitedefaultendpunct}{\mcitedefaultseppunct}\relax
\EndOfBibitem
\bibitem[Rubinstein and Colby(2003)]{rubincolby}
M.~Rubinstein and R.~Colby, \emph{Polymer Physics}, Oxford University Press,
  2003\relax
\mciteBstWouldAddEndPuncttrue
\mciteSetBstMidEndSepPunct{\mcitedefaultmidpunct}
{\mcitedefaultendpunct}{\mcitedefaultseppunct}\relax
\EndOfBibitem
\bibitem[Doi and Edwards(1986)]{doied}
M.~Doi and S.~Edwards, \emph{The {T}heory of {P}olymer {D}ynamics}, Oxford
  University Press, 1986\relax
\mciteBstWouldAddEndPuncttrue
\mciteSetBstMidEndSepPunct{\mcitedefaultmidpunct}
{\mcitedefaultendpunct}{\mcitedefaultseppunct}\relax
\EndOfBibitem
\bibitem[Mc{L}eish(2002)]{mcleish02AiP}
T.~Mc{L}eish, \emph{Advances in Physics}, 2002, \textbf{51}, 1379\relax
\mciteBstWouldAddEndPuncttrue
\mciteSetBstMidEndSepPunct{\mcitedefaultmidpunct}
{\mcitedefaultendpunct}{\mcitedefaultseppunct}\relax
\EndOfBibitem
\bibitem[Colby \emph{et~al.}(1987)Colby, Fetters, and Graessey]{colby87MAC}
R.~Colby, L.~Fetters and W.~Graessey, \emph{Macromolecules}, 1987, \textbf{20},
  2226\relax
\mciteBstWouldAddEndPuncttrue
\mciteSetBstMidEndSepPunct{\mcitedefaultmidpunct}
{\mcitedefaultendpunct}{\mcitedefaultseppunct}\relax
\EndOfBibitem
\bibitem[Dee and Sauer(1998)]{dee98AiP}
G.~Dee and B.~Sauer, \emph{Advances in Physics}, 1998, \textbf{47}, 161\relax
\mciteBstWouldAddEndPuncttrue
\mciteSetBstMidEndSepPunct{\mcitedefaultmidpunct}
{\mcitedefaultendpunct}{\mcitedefaultseppunct}\relax
\EndOfBibitem
\bibitem[Wu(1970)]{wu70JPC}
S.~Wu, \emph{Journal of Physical Chemistry}, 1970, \textbf{74}, 632\relax
\mciteBstWouldAddEndPuncttrue
\mciteSetBstMidEndSepPunct{\mcitedefaultmidpunct}
{\mcitedefaultendpunct}{\mcitedefaultseppunct}\relax
\EndOfBibitem
\bibitem[Fetters \emph{et~al.}(1999)Fetters, Lohse, and Milner]{fetters3}
L.~Fetters, D.~Lohse and S.~Milner, \emph{Macromolecules}, 1999, \textbf{32},
  6847\relax
\mciteBstWouldAddEndPuncttrue
\mciteSetBstMidEndSepPunct{\mcitedefaultmidpunct}
{\mcitedefaultendpunct}{\mcitedefaultseppunct}\relax
\EndOfBibitem
\bibitem[Bach \emph{et~al.}(2003)Bach, Almdal, Rasmussen, and
  Hassager]{bach2003}
A.~Bach, K.~Almdal, H.~K. Rasmussen and O.~Hassager, \emph{Macromolecules},
  2003, \textbf{36}, 5174\relax
\mciteBstWouldAddEndPuncttrue
\mciteSetBstMidEndSepPunct{\mcitedefaultmidpunct}
{\mcitedefaultendpunct}{\mcitedefaultseppunct}\relax
\EndOfBibitem
\bibitem[Williams \emph{et~al.}(1955)Williams, Landel, and
  Ferry]{williams55JACS}
M.~Williams, R.~Landel and J.~Ferry, \emph{Journal of the American Chemical
  Society}, 1955, \textbf{77}, 3701\relax
\mciteBstWouldAddEndPuncttrue
\mciteSetBstMidEndSepPunct{\mcitedefaultmidpunct}
{\mcitedefaultendpunct}{\mcitedefaultseppunct}\relax
\EndOfBibitem
\end{mcitethebibliography}
\providecommand*{\mcitethebibliography}{\thebibliography}
\csname @ifundefined\endcsname{endmcitethebibliography}
{\let\endmcitethebibliography\endthebibliography}{}

}
\end{document}